\documentclass[AMA,STIX2COL]{MRM}
\articletype{Technical Note}%
\usepackage{amsmath,amssymb,bm}
\newcommand{\Dmax}{\mathrm{D}_{\max}}
\newcommand{\Dzero}{\mathrm{D}_0}
\newcommand{\Dtwo}{\mathrm{D}_2}
\newcommand{\Dtwom}{\mathrm{D}^{2m}}

\newcommand{\Qzero}{\mathrm{Q}_0}
\newcommand{\Qtwom}{\mathrm{Q}^{2m}}

\newcommand{\Tzero}{\mathrm{T}_0}
\newcommand{\Ttwom}{\mathrm{T}^{2m}}

\newcommand{\E}{\mathbb{E}}
\newcommand{\Var}{\operatorname{Var}}
\newcommand{\Cov}{\operatorname{Cov}}

\newcommand{\sphere}{S^2}
\newcommand{\ubar}{\mathbf{u}}
\newcommand{\Dmicro}{\bm{\mathcal{D}}}
\newcommand{\Dmean}{\bm{\mathrm{D}}}
\newcommand{\Dmicrozero}{\mathcal{D}_0}
\newcommand{\Dmicrotwo}{\mathcal{D}_2}
\newcommand{\Dmicrotwom}{\mathcal{D}^{2m}}
\newcommand{\Dmicrotwomstar}{\mathcal{D}^{2m*}}
\newcommand{\Ibold}{\bm{\mathrm{I}}}
\newcommand{\Dtwomstar}{\mathrm{D}^{2m*}}

\begin{document}

\makeatletter
\renewcommand{\@@address}[2][]{%
\g@addto@macro\jmkaddress{\hsize=\textwidth%
\def\baselinestretch{1}%
\stepcounter{affn}%
\xdef\@currentlabel{\theaffn}%
\jmkLabel{#1}%
\addressfont\if\relax\detokenize{#1}\relax\else
\@hangfrom{\textsuperscript{\smash{\theaffn}}}\fi%
#2\vphantom{Thy}\par%
}}%
\makeatother

\title{Constrained estimation of rotational invariants of the 
cumulant expansion (RICE) for rapid tensor-valued diffusion MRI}

\author[1,2,3]{Jinyang Yu*}{}
\author[1,3,4]{Oliver G\"odicke*}{}
\author[5]{Frederik B. Laun}{}
\author[6,7]{Obada T. Alhalabi}{}
\author[3,8]{Iris A. Kohler}{}
\author[3,8,9,10]{J\"urgen Hesser}{}
\author[6,7]{Sandro M. Krieg}{}
\author[6,7]{Bogdana Suchorska}{}
\author[3, 11]{Heinz-Peter Schlemmer}{}
\author[1,3,4,6]{Mark E. Ladd}{}
\author[3, 6,11,12]{David Bonekamp}{}
\author[11]{Johann M. E. Jende}{}
\author[1,3,4]{Tristan A. Kuder}{}

\authormark{YU \& G\"ODICKE \textit{et al.}}

\address[1]{
\orgdiv{Division of Medical Physics in Radiology},
\orgname{German Cancer Research Center (DKFZ)},
\orgaddress{\city{Heidelberg}, \country{Germany}}
}
\address[2]{
\orgdiv{Faculty of Engineering Sciences},
\orgname{Heidelberg University},
\orgaddress{\city{Heidelberg}, \country{Germany}}
}
\address[3]{
\orgdiv{CZS Heidelberg Initiative for Model-Based AI (MBAI)},
\orgname{Heidelberg University},
\orgaddress{\city{Heidelberg}, \country{Germany}}
}
\address[4]{
\orgdiv{Faculty of Physics and Astronomy},
\orgname{Heidelberg University},
\orgaddress{\city{Heidelberg}, \country{Germany}}
}
\address[5]{
\orgdiv{Institute of Radiology},
\orgname{Uniklinikum Erlangen, Friedrich-Alexander-Universit\"at
Erlangen-N\"urnberg (FAU)},
\orgaddress{\city{Erlangen}, \country{Germany}}
}
\address[6]{
\orgdiv{Faculty of Medicine},
\orgname{Heidelberg University},
\orgaddress{\city{Heidelberg}, \country{Germany}}
}
\address[7]{
\orgdiv{Department of Neurosurgery},
\orgname{University Hospital Heidelberg},
\orgaddress{\city{Heidelberg}, \country{Germany}}
}
\address[8]{
\orgdiv{Mannheim Institute for Intelligent Systems in Medicine (MIISM)},
\orgname{Heidelberg University},
\orgaddress{\city{Mannheim}, \country{Germany}}
}
\address[9]{
\orgdiv{Interdisciplinary Center for Scientific Computing (IWR)},
\orgname{Heidelberg University},
\orgaddress{\city{Heidelberg}, \country{Germany}}
}
\address[10]{
\orgdiv{Institute for Computer Engineering (ZITI)},
\orgname{Heidelberg University},
\orgaddress{\city{Heidelberg}, \country{Germany}}
}
\address[11]{
\orgdiv{Division of Radiology},
\orgname{German Cancer Research Center (DKFZ)},
\orgaddress{\city{Heidelberg}, \country{Germany}}
}
\address[12]{
\orgname{National Center for Tumor Diseases (NCT) Heidelberg},
\orgaddress{\city{Heidelberg}, \country{Germany}}
}

\corres{Tristan A. Kuder\\
Im Neuenheimer Feld 280, 69120 Heidelberg, Germany\\
\email{t.kuder@dkfz-heidelberg.de}\\
*Jinyang Yu and Oliver G\"odicke contributed equally to this work.}

\finfo{}
\doiheadtext{}
\doi{}
\historydates{}
\gdef\printjnlcitation{}

\abstract[ABSTRACT]{
\section{Purpose}
To complement 1.5-minute measurements of common tensor-valued diffusion MRI (dMRI) markers with rapid constrained fitting.

\section{Methods} 
Fast dMRI protocols for obtaining rotational invariants of the cumulant expansion (RICE) were paired with constrained weighted linear least squares (CWLLS) to stabilize the more fragile WLLS fit. A compact constraint set was formulated, including a novel mean-dependent upper bound on total diffusional variance. Evaluation used diffusion tensor distribution (DTD) simulations, healthy-volunteer data with a resolution-dependent SNR experiment, and a glioma patient dataset. A 5-minute q-space trajectory imaging (QTI) protocol served as a reference.

\section{Results} 
Across experiments, CWLLS reduced unphysical estimates and fit outliers in parameters such as microscopic FA and isotropic diffusivity variance. In simulations, it narrowed error distributions most clearly in the CSF-dominant case, while some metrics showed a bias--variance trade-off. In vivo, CWLLS removed negative variance estimates, truncated out-of-bounds tails, and reduced artifacts in fluid-contaminated voxels while preserving anatomical contrast. It also retained more stable maps than WLLS at higher resolution, although both estimators degraded in the lowest-SNR setting. Notably, the new mean-dependent variance bound was violated in 15.4\% of voxels in the patient dataset, accounting for nearly half of the 32.7\% that violated at least one constraint. Healthy-volunteer benchmarking showed that CWLLS completed in under 30 seconds. The constrained QTI fit required 72 minutes, making CWLLS 160 times faster.

\section{Conclusion} 
CWLLS for fast RICE yielded high-quality parameter maps at an online-ready computational cost. This may enhance the reliability of dMRI tissue characterization and strengthen the path toward clinical translation.

\section{Keywords} 
diffusion MRI; tensor-valued diffusion encoding; rotational invariants; constrained estimation; q-space trajectory imaging
}

\jnlcitation{}

\makeatletter
\patchcmd{\@maketitle}{\removelastskip\vspace*{20pt}}{\removelastskip\vspace*{3pt}}{}{}
\patchcmd{\@maketitle}{\removelastskip\vskip22pt}{\removelastskip\vskip10pt}{}{}
\patchcmd{\@maketitle}{\removelastskip\vskip18pt}{\removelastskip\vskip10pt}{}{}
\patchcmd{\@maketitle}{\removelastskip\vskip19pt}{\removelastskip\vskip8pt}{}{}
\patchcmd{\@abstract}{\vspace*{8.5\p@}}{\vspace*{4\p@}}{}{}
\patchcmd{\@abstract}{\baselineskip15pt}{\baselineskip10.5pt}{}{}
\patchcmd{\printabstractpart}{\removelastskip\vskip15pt}{\removelastskip\vskip6pt}{}{}
\patchcmd{\printabstractpart}{\removelastskip\vskip15pt}{\removelastskip\vskip6pt}{}{}
\def\titlefont{\rmfamily\fontsize{17.28}{19}\bfseries\selectfont\leftskip\z@\rightskip\z@ plus1fil\let\mathbcal\titmathbcal}
\def\Authorfont{\rmfamily\fontsize{10}{13}\selectfont\leftskip\z@\rightskip\z@ plus1fil}
\def\addressfont{\hsize\abs@coli@hsize\rmfamily\fontsize{7}{8}\selectfont\leftskip\z@\rightskip\z@ plus1fil}
\def\corresfont{\hsize\abs@coli@hsize\rmfamily\fontsize{7}{8.5}\selectfont\leftskip\z@\rightskip\z@ plus1fil}
\def\abstractfont{\hsize\abs@colii@hsize\rmfamily\fontsize{8}{10.5}\selectfont\leftskip7\p@\rightskip\leftskip}
\def\absheadfont{\hsize\abs@colii@hsize\rmfamily\fontsize{8}{9}\bfseries\selectfont\leftskip7\p@\rightskip\leftskip}
\adjtitleskip=0pt
\makeatother

\maketitle

\section{Introduction}

Tensor-valued diffusion encoding varies the shape of the $b$-tensor to separate sources of diffusional variance that conventional linear encoding cannot distinguish \cite{westin2016qspace}. Q-space trajectory imaging (QTI) combines free-waveform encoding with a diffusion tensor distribution (DTD) model and yields parameters such as microscopic fractional anisotropy ($\mu$FA), which has shown disease- and treatment-related changes in brain tumors, white-matter injury, breast cancer, and prostate cancer \cite{nilsson2020intracranial,zhou2024stroke,boito2023covid,cho2024breast,langbein2021prostate,lerner2026radiotherapy}.

This information gain comes at a cost. QTI estimates the first two DTD cumulants: the mean diffusion tensor and its fourth-order covariance, which contain six and 21 independent elements, respectively. Since the covariance contribution enters the signal through the weaker $b^2$ term, stable estimation requires lengthy measurements with many diffusion encodings and sufficient SNR. Shorter scans and smaller voxels therefore make parameter estimation increasingly fragile: the standard weighted linear least squares (WLLS) fit is fast, but sparse or low-SNR data can yield unphysical parameter values and visible fitting artifacts. Together, these challenges complicate broader clinical adoption.

The rotational invariants of the cumulant expansion (RICE) framework by Coelho et al. offers a solution to the acquisition side of this problem: RICE reformulates the same DTD cumulant model using SO(3)-irreducible representations~\cite{coelho2026geometry}. The diffusion and covariance tensors are decomposed into orthogonal sectors of angular degree $\ell=0$, $\ell=2$, and $\ell=4$. With adequate directional sampling, this allows the most common DTD-derived markers, such as $\mu$FA, mean kurtosis (MK), and isotropic diffusivity variance ($V_{\mathrm{MD}}$), to be expressed through a small subset of rotationally invariant quantities. This compact representation motivates instant RICE (iRICE) protocols, which target only the information required for selected contrasts and therefore require far fewer measurements than full QTI tensor estimation fitting 27 independent tensor components~\cite{coelho2026geometry}. In practice, iRICE reduces whole-brain acquisitions to one to two minutes. However, it also leaves less redundancy to stabilize the WLLS fit.

The estimator side of the problem therefore remains. For the full QTI cumulant fit, Herberthson et al. introduced QTI+, which enforces positive semidefiniteness of the mean, covariance, and second-moment tensors \cite{herberthson2021qti}. Boito et al. extended this framework with upper diffusivity bounds in QTI$\pm$ \cite{boito2023diffusivity}. These constraints improved robustness for shorter protocols and lower SNR, particularly in free-water- and CSF-rich voxels where unconstrained fits can yield physically infeasible values. This gain comes at a computational cost, as semidefinite optimization over all cumulant tensor components can take several tens of minutes per dataset on a standard workstation. Nevertheless, these studies establish physical constraints as an essential criterion alongside residual error.

Existing full QTI constraints cannot be transferred to iRICE by a simple change of basis. They act on the complete QTI parameter space, including covariance information that is discarded in an iRICE acquisition. iRICE estimates only the irreducible components required for the targeted invariant maps \cite{coelho2026geometry}. A constrained iRICE estimator must therefore act directly on this reduced parameterization.

In this work, we introduce constrained weighted linear least squares (CWLLS) for iRICE. Our method formulates a compact constraint set directly on the fitted iRICE coefficients. It bounds the reconstructed mean tensor, enforces non-negative scalar variance terms, and adds a new, tighter mean-dependent limit on total diffusivity variance. To retain fitting speed, CWLLS solves only voxels whose initial estimates approach or violate a constraint bound. We show that these constraints bring significant stability gains to iRICE while preserving its speed advantage from acquisition through parameter estimation.

We evaluated CWLLS in DTD simulations and healthy-volunteer and glioma data. In vivo, we compared CWLLS and WLLS using QTI$\pm$ fits from a five-minute regular protocol as reference. We also challenged both iRICE estimators with progressively higher spatial resolution and lower SNR. Across these experiments, CWLLS reduced unphysical estimates and fit outliers. Selective constrained optimization completed in under 30 seconds, 160-fold faster than QTI$\pm$, yielding high-quality parameter maps from a 1.5-minute acquisition and unlocking the potential of online clinical fitting.

\section{Methods}

\subsection{Signal representation}

For a voxel represented by a distribution of Gaussian diffusion compartments, the tensor-valued diffusion MRI signal can be written as cumulant expansion to second order~\cite{westin2016qspace,vankampen1981stochastic,kiselev2010cumulant},
\begin{equation}
\ln\frac{S(\mathrm{\mathbf{B}})}{S_0}
=-\mathrm{B}_{ij}\mathrm{D}_{ij}
+\frac{1}{2}\mathrm{B}_{ij}\mathrm{B}_{kl}\mathrm{C}_{ijkl}
+O(b^3),
\label{eq:signal_model}
\end{equation}
where $S_0$ is the non-diffusion-weighted signal, $\mathrm{D}_{ij}=\langle
\mathcal{D}_{ij}\rangle$ is the voxel-wise mean diffusion tensor,
$\mathrm{C}_{ijkl}=\langle(\mathcal{D}_{ij}-\langle
\mathcal{D}_{ij}\rangle)(\mathcal{D}_{kl}-\langle
\mathcal{D}_{kl}\rangle)\rangle$ is the DTD covariance tensor, and $\mathrm{B}_{ij}$ is the
measurement $b$-tensor.

RICE decomposes $\mathrm{D}_{ij}$ and $\mathrm{C}_{ijkl}$ into SO(3)-irreducible spherical tensor components, such that Eq.~\ref{eq:signal_model} contains $\ell=0$, $\ell=2$, and $\ell=4$ sectors \cite{coelho2026geometry}. For axially symmetric encoding with shape parameter $\beta$~\cite{szczepankiewicz2021gradient}, the acquisition uses linear tensor encoding (LTE, $\beta=1$) and spherical tensor encoding (STE, $\beta=0$). iRICE retains only the lower-degree terms needed for the targeted invariants and does not explicitly fit the $\ell=4$ sector:

\begin{align}
&\ln\frac{S(b,\hat{\mathbf{g}})}{S|_{b=0}}
=
\mathcal{L}_0+\mathcal{L}_2,
\label{eq:rice_irice_signal}\\
&\mathcal{L}_0
=
-b\Dzero
+\frac{b^2}{2}
\left(
\Qzero+\beta^2\Tzero
\right),
\nonumber\\
&\mathcal{L}_2
=
\sum_{m=-2}^{2}
\left[
-b\beta\Dtwom
+\frac{b^2}{2}
\left(
\beta\Qtwom
+\beta^2\Ttwom
\right)
\right]
Y^{2m}(\hat{\mathbf{g}}).
\nonumber
\end{align}
Here, $Y^{2m}(\hat{\mathbf{g}})$ are the Racah-normalized $\ell=2$ spherical harmonics. $\Dzero$ and $\Dtwom$ denote the $\ell=0$ and $\ell=2$ mean-tensor components, and $\Qzero$ and $\Tzero$ denote the scalar size- and shape-related covariance terms. iRICE resolves these eight degrees of freedom while discarding the $\Qtwom$ and $\Ttwom$ components~\cite{coelho2026geometry}.

For both constraint derivation and metric computation, $\Dtwom$ enters through its squared norm,
\begin{equation}
\Dtwo^2=\sum_{m=-2}^{2}\Dtwom{\Dtwom}^{*}.
\label{eq:d2_norm}
\end{equation}
After fitting Eq.~\ref{eq:rice_irice_signal}, MD, FA, MK, $\mu$FA, $V_{\mathrm{MD}}$ and the corresponding normalized metric ($C_{\mathrm{MD}}$)~\cite{westin2016qspace} were computed from the eight fitted iRICE coefficients, as given in the Supporting Information.

\subsection{WLLS estimator}

WLLS served as the baseline log-signal estimator, with predicted-signal weights accounting for heteroskedasticity from the log transformation \cite{coelho2026geometry,herberthson2021qti,veraart2013wlls}. For voxel \(v\), let \(\mathbf{A}\) denote the iRICE design matrix, \(\mathbf{y}_v\) the measured log-signal vector, and \(\boldsymbol{\theta}_v\) the fitted iRICE coefficient vector. From an initial ordinary least-squares estimate \(\hat{\boldsymbol{\theta}}^{\mathrm{OLS}}_v\), the predicted signal was computed as
\begin{equation}
\hat{\mathbf{s}}_v
=
\exp(\mathbf{A}\hat{\boldsymbol{\theta}}^{\mathrm{OLS}}_v),
\end{equation}
and the square-root weight matrix was then defined as
\begin{equation}
\mathbf{W}_v
=
\mathrm{diag}(\hat{\mathbf{s}}_v).
\end{equation}
The WLLS estimate then used these fixed weights in the optimization
\begin{equation}
\hat{\boldsymbol{\theta}}_v
=
\arg\min_{\boldsymbol{\theta}}
\left\|
\mathbf{W}_v
\left(
\mathbf{A}\boldsymbol{\theta}-\mathbf{y}_v
\right)
\right\|_2^2.
\label{eq:wlls_objective}
\end{equation}

\subsection{CWLLS estimator}
CWLLS used the same weighted objective as WLLS, subject to the mean-tensor, scalar non-negativity, and mean-dependent variance constraints in Eqs.~\ref{eq:mean_tensor_constraint}--\ref{eq:mean_dependent_bound}. The reconstructed mean diffusion tensor was constrained in the Cartesian basis by
\begin{equation}
0\preceq \bm{\mathcal{D}}(\boldsymbol{\theta}) \preceq \Dmax \mathbf{I},
\label{eq:mean_tensor_constraint}
\end{equation}
where $\Dmax=3.075~\mu\mathrm{m}^2/\mathrm{ms}$ \cite{boito2023diffusivity}. The scalar variance terms satisfied the non-negativity conditions
\begin{equation}
\Qzero\ge0,
\qquad
\Tzero\ge0.
\label{eq:scalar_nonnegativity}
\end{equation}
Finally, we derived and imposed the mean-dependent diffusivity-limited variance bound
\begin{equation}
\Qzero
+\Tzero
+\frac{1}{5}\Dtwo^2
\le
\Dzero
\left[
\Dmax-\Dzero
\right].
\label{eq:mean_dependent_bound}
\end{equation}
This follows from bounded microscopic directional diffusivities after rotational averaging. It is the key iRICE-specific constraint and provides a tighter, mean-dependent alternative to the constant scalar cap motivated by diffusivity-limited QTI$\pm$ \cite{boito2023diffusivity}. For post hoc diagnostic comparison only, we also evaluated the constant-cap form \cite{Yu2026_26327d5e}
\begin{equation}
\Qzero+\Tzero+\frac{1}{5}\Dtwo^2
\le
\frac{\Dmax^2}{4}.
\label{eq:constant_variance_cap}
\end{equation}
Their derivation and scope are provided in the Supporting Information.

All CWLLS constraints define a convex feasible set in the fitted parameters. Optimization was implemented in CVXPY \cite{diamond2016cvxpy,agrawal2018rewriting}, with MOSEK \cite{mosek2025optimizer} as the numerical solver.

\subsection{Constraint screening and violation assessment}

Constraint violations were assessed for WLLS and final CWLLS estimates within the fitting mask. To exclude numerical tolerance effects, an estimate was classified as a violation only if it exceeded the allowed range by more than an absolute tolerance of $10^{-7}$. The same tolerance was used pre- and post-fit.

To reduce computational cost, only WLLS estimates $\boldsymbol{\theta}_v$ violating at least one bound underwent constrained optimization, all others were retained.

\subsection{DTD simulation}

We performed single-voxel DTD simulations using two discrete tensor distributions. Following prior DTD simulation work \cite{reymbaut2020accuracy,topgaard2019dtd}, each axisymmetric tensor was written as
\begin{equation}
\bm{\mathcal{D}}
=
\lambda_{\perp}\mathbf{I}
+
(\lambda_{\parallel}-\lambda_{\perp})
\mathbf{u}\mathbf{u}^{\mathrm{T}},
\end{equation}
where $\mathbf{u}$ is the symmetry-axis direction, and $\lambda_{\parallel}$ and $\lambda_{\perp}$ are the parallel and perpendicular diffusivities, respectively.

Two DTD configurations were simulated. The first represented a CSF-dominant high-diffusivity mixture, combining a fast isotropic tensor with a low fraction of a white-matter-like prolate population. The second represented a white-matter-like configuration composed of prolate tensors with moderate angular dispersion. The simulation recipes are shown in Figure~\ref{fig:simulation}.

Noise-free signals were generated from Eq.~\ref{eq:signal_model} using the theoretical minimal 14-encoding iRICE protocol: one $b=0$ image, six LTE directions at $b=1.0~\mathrm{ms}/\mu\mathrm{m}^2$, six LTE directions at $b=2.0~\mathrm{ms}/\mu\mathrm{m}^2$, and one STE measurement at $b=2.0~\mathrm{ms}/\mu\mathrm{m}^2$. The LTE shells used icosahedral vertices to obtain six antipodally distinct directions, following the spherical 4-design in \cite{coelho2026geometry}. Rician magnitude noise was added at SNR~30 relative to $S_0$ \cite{reymbaut2020accuracy,gudbjartsson1995rician}. For each DTD, 1000 independent noise realizations were fitted with WLLS and CWLLS. Deviations were computed as fitted scalar values minus DTD ground-truth invariants.

\subsection{In vivo measurements and evaluation}

In vivo brain data were acquired from 14 healthy subjects and one glioma patient on a 3T Prisma Fit scanner (Siemens Healthineers, Erlangen, Germany) using a custom free-waveform sequence \cite{martin2020cnr,goedicke2024accelerated} and a 64-channel head coil. The study was approved by the local institutional review board, and all participants provided written informed consent (IRB approvals S-184/2018 and S-788/2024). The patient scan was conducted before treatment and included the standard preoperative MRI protocol.

The necessary acquisition time for the QTI reference data was 5 min and used 54 encodings across LTE, PTE, and STE shapes. Imaging parameters were 2.5~mm isotropic voxel size, matrix size $88\times88\times24$, repetition time 4400~ms, echo time 84~ms, receiver bandwidth 1496~Hz/pixel, iPAT 2, and partial Fourier factor $6/8$ \cite{goedicke2024accelerated,goedicke2025inversion}. The iRICE subset had a nominal acquisition time of 1.5 min and included 20 encodings: one $b=0$ image, 10 LTE directions at $b=1.0~\mathrm{ms}/\mu\mathrm{m}^2$, six LTE directions at $b=2.0~\mathrm{ms}/\mu\mathrm{m}^2$, and three STE measurements at $b=2.0~\mathrm{ms}/\mu\mathrm{m}^2$. The six-direction LTE set used the same icosahedral vertices as in the simulations, whereas the ten-direction LTE set was constructed from antipodally distinct vertices of a dodecahedron, again following the spherical-design sampling scheme \cite{coelho2026geometry}.

An additional healthy volunteer was scanned using isotropic voxel sizes from 2.5 to 1.5~mm to assess estimator behavior with decreasing voxel volume and lower effective SNR. For the patient scan, the repetition time was increased to 6000~ms and the resolution was changed to $2.0\times2.0\times4.0~\mathrm{mm}^3$, resulting in minor changes to the echo time and receiver bandwidth.

Diffusion data were denoised and corrected for Rician bias, Gibbs ringing, susceptibility distortion, and motion using established diffusion MRI preprocessing methods \cite{veraart2016denoising,tournier2019mrtrix3, andersson2003susceptibility,klein2010elastix,nilsson2015extrapolation}. Regular QTI data were fitted with QTI$\pm$ \cite{boito2023diffusivity} as an in vivo reference. iRICE data were fitted with both WLLS and CWLLS. All estimator comparisons used the same preprocessing outputs and analysis masks and were run using the same computational resources: an Intel Core i9-13900 processor (2.00~GHz, 24 cores).

We compared healthy-volunteer fits qualitatively and with difference maps. For the glioma patient dataset, we assessed constraint-violation overlays and whole-brain histograms of the fitted parameter values.

\section{Results}

\subsection{DTD simulation}
\begin{figure*}[!t]
\centering
\includegraphics[width=0.90\textwidth]{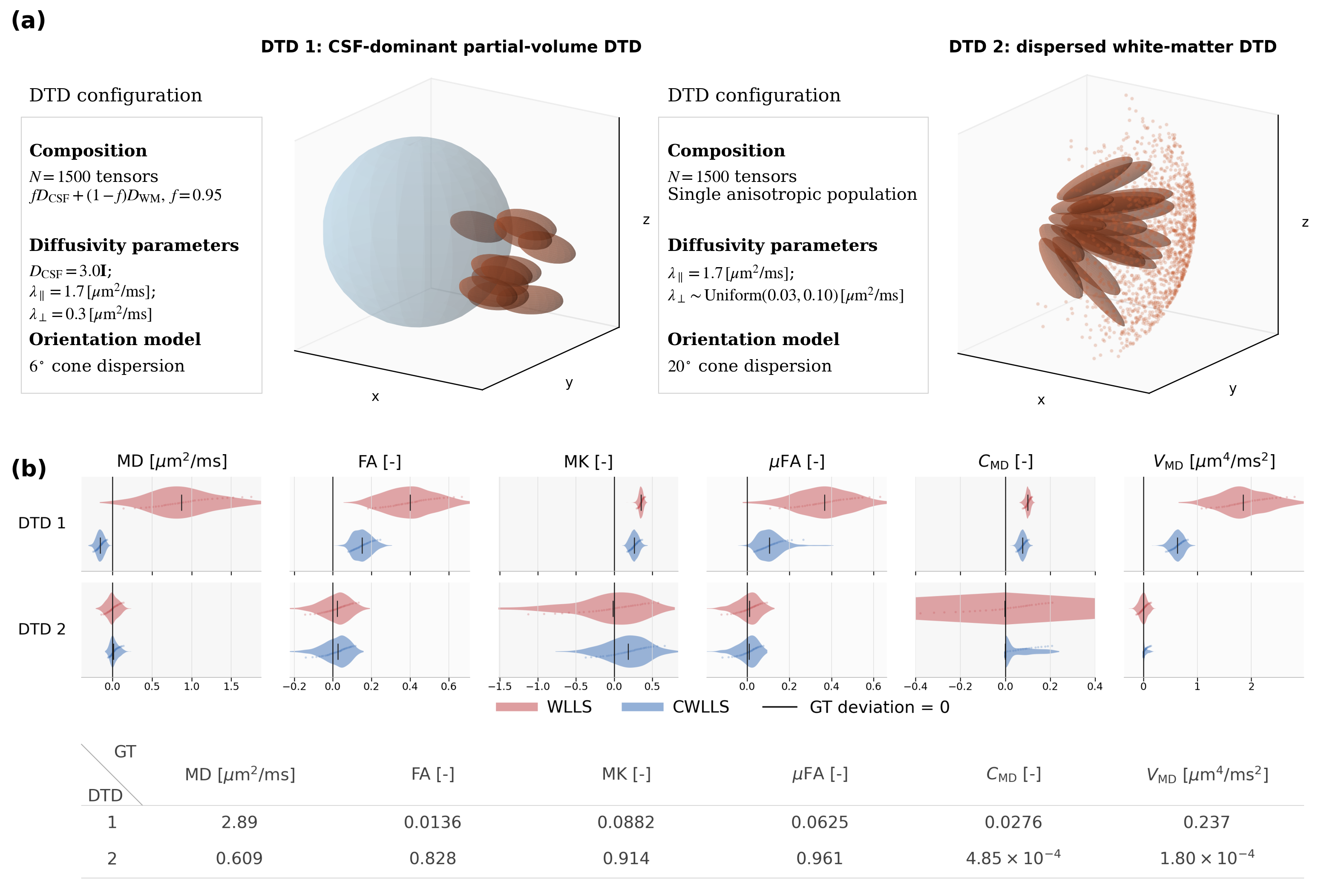}
\caption{
DTD simulation at SNR 30. Synthetic iRICE signals were corrupted with Rician noise and fitted with
WLLS and CWLLS. (a) DTD configurations and visualized tensor distributions for DTD 1, a
CSF-dominant partial-volume DTD, and DTD 2, a dispersed white-matter DTD. 
The listed tensor eigenvalues and $D_{\mathrm{CSF}}$ and $D_{\mathrm{WM}}$ values use the same units as MD.
(b) Estimator deviations,
defined as fitted values minus the corresponding DTD ground truth, for MD, FA, MK, $\mu$FA,
$C_{\mathrm{MD}}$, and $V_{\mathrm{MD}}$. Rows correspond to DTD 1 and DTD 2. 
The table below reports the corresponding absolute DTD ground-truth values for all six metrics. The vertical line
marks zero deviation, and short black lines indicate medians. In the CSF-dominant DTD 1, CWLLS
yields substantially more accurate parameter estimates with narrower distributions, i.e., lower
variance, than unconstrained WLLS. In the dispersed white-matter DTD 2, CWLLS generally reduces
the dispersion of the estimates. An exception is MK, for which the narrower distribution is
accompanied by increased bias.
}
\label{fig:simulation}
\end{figure*}

Figure~\ref{fig:simulation}(b) shows estimator deviations from DTD ground truth at SNR 30. In the CSF-dominant DTD, WLLS produced broad positive deviations and long tails in several metrics, including MD, FA, $\mu$FA, and $V_{\mathrm{MD}}$. CWLLS narrowed these distributions, reducing extreme deviations and yielding more accurate estimates. The MD distribution was pulled back toward the admissible diffusivity range, consistent with the mean-tensor upper bound.

In the dispersed white-matter DTD, estimator differences were smaller for MD and FA, indicating limited constraint effects on the primary diffusion-tensor contrasts. Differences were more visible in MK, $C_{\mathrm{MD}}$, and $V_{\mathrm{MD}}$, where the constraints again reduced the spread of distributions. While CWLLS reduced some WLLS tails in MK, it also shifted the median toward larger errors. The simulation therefore points to a robustness benefit rather than uniform accuracy gains: CWLLS suppresses infeasible or extreme estimates, but some metrics can still shift away from the ground truth.

\subsection{Healthy volunteer}
\label{subsec:healthy_volunteer}
\begin{figure*}[!t]
\centering
\includegraphics[width=0.90\textwidth]{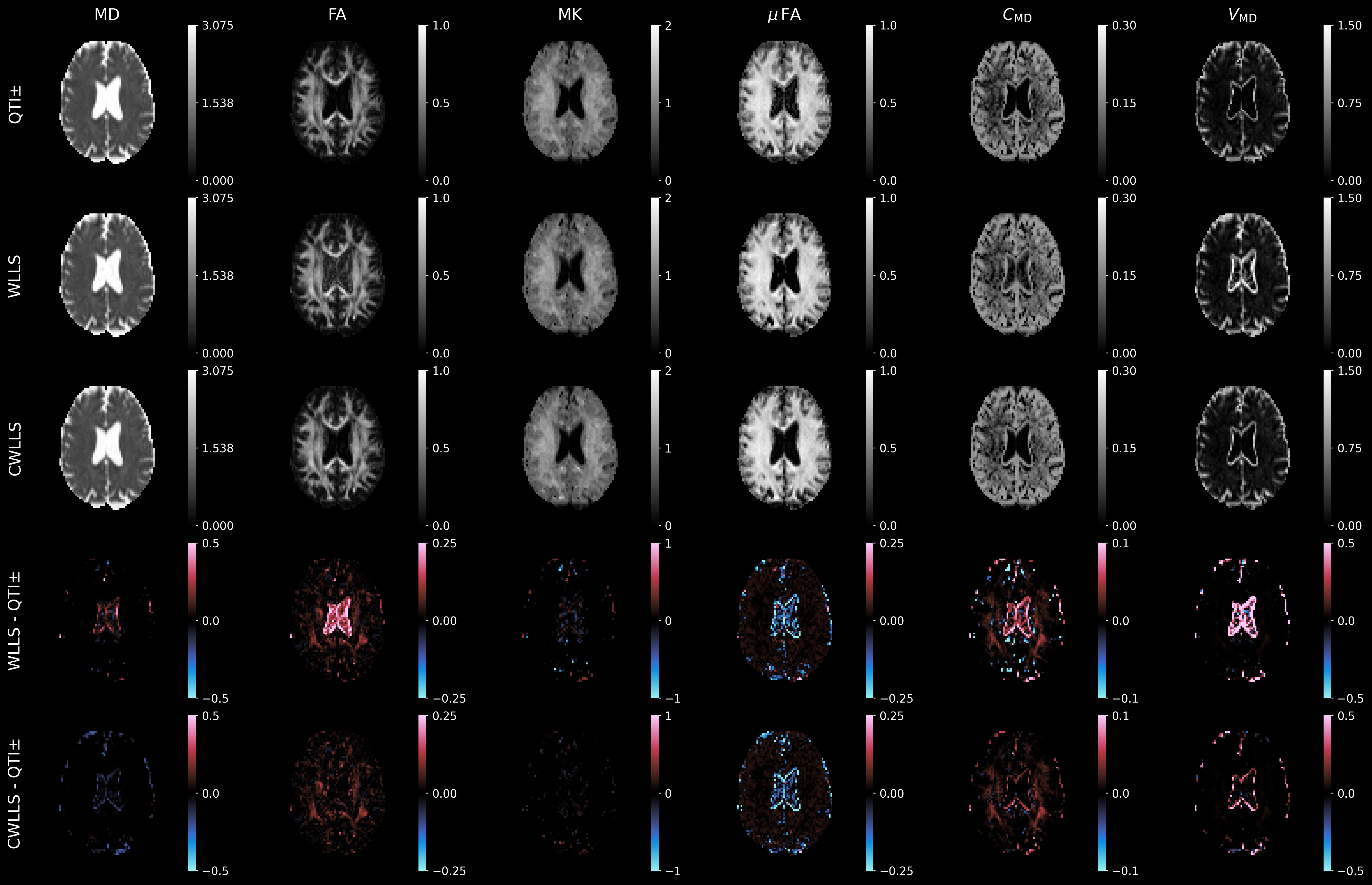}
\caption{
Healthy volunteer estimator comparison using the same regular 54-encoding protocol for all three
fits. Maps are shown for MD, FA, MK, $\mu$FA, $C_{\mathrm{MD}}$, and
$V_{\mathrm{MD}}$. The first row shows QTI$\pm$ and is used as an in vivo reference. The second
and third rows show WLLS and CWLLS fits, respectively. The fourth and fifth rows show the
corresponding WLLS--QTI$\pm$ and CWLLS--QTI$\pm$ difference maps. Unconstrained WLLS produces the
largest number of outliers in fluid-rich voxels, particularly in the ventricles and at cortical
CSF interfaces. The QTI$\pm$ reference largely suppresses these outliers. CWLLS achieves similar
visual quality on the same data, although isolated outliers remain in MD, FA, MK, and $\mu$FA.
The isotropic variance-related parameters $C_{\mathrm{MD}}$ and $V_{\mathrm{MD}}$ show a distinct
pattern: constrained fitting yields lower values at tissue--fluid interfaces, where CSF partial
volume can produce high isotropic variance and may coincide with constraint violations. This
effect is most apparent in the high-valued $C_{\mathrm{MD}}$/$V_{\mathrm{MD}}$ rim around the
ventricles, which appears thinner and lower-valued in the constrained fits. MD is reported in
$\mu\mathrm{m}^{2}/\mathrm{ms}$, and $V_{\mathrm{MD}}$ is reported in
$\mu\mathrm{m}^{4}/\mathrm{ms}^{2}$, while all other metrics are dimensionless.
}
\label{fig:healthy_maps}
\end{figure*}

Figure~\ref{fig:healthy_maps} shows maps from a healthy volunteer. For a fair estimator comparison, QTI$\pm$, WLLS, and CWLLS were applied to the same 54-encoding protocol. The WLLS and CWLLS estimates preserved the main anatomical contrast across MD, FA, MK, $\mu$FA, $C_{\mathrm{MD}}$, and $V_{\mathrm{MD}}$.

Estimator-dependent differences were most visible in fluid-rich regions, particularly in the ventricles and at cortical CSF interfaces. WLLS produced the largest number of outlying voxels, whereas CWLLS reduced these artifacts and approached the visual quality of the QTI$\pm$ reference. In $C_{\mathrm{MD}}$ and $V_{\mathrm{MD}}$, constrained fitting thinned and lowered the high-valued rim around the ventricles, consistent with reduced spuriously high isotropic variance in CSF partial volume voxels. Additional 20-encoding iRICE maps and cohort-level invariant distributions across 14 healthy volunteers are shown in Supporting Figures S1 and S2, respectively.

\subsection{Glioma patient}

\begin{figure*}[!t]
\centering
\includegraphics[width=0.90\textwidth]{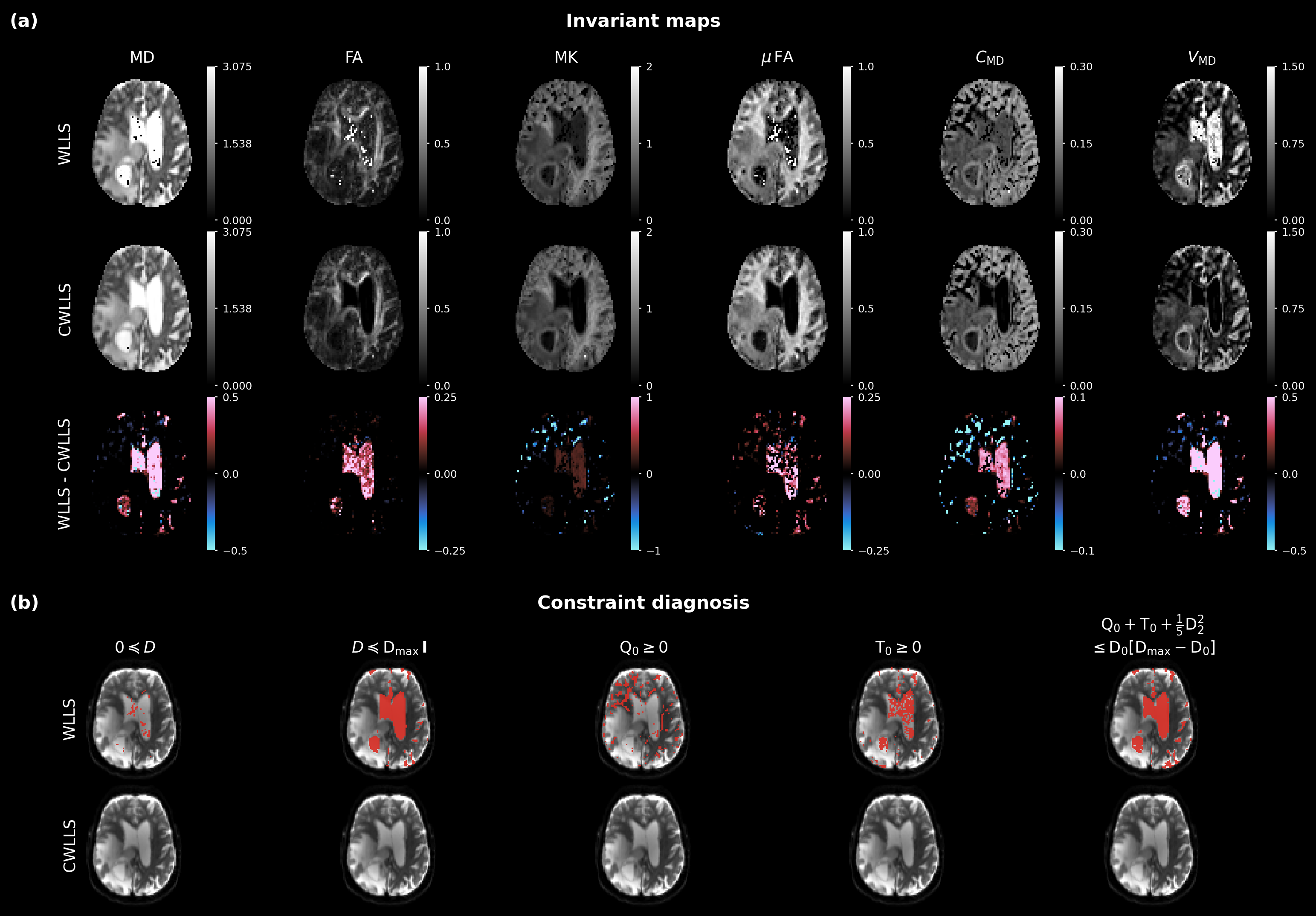}
\caption{
Estimator comparison and constraint diagnosis in a glioma patient acquired with the
1.5-min, 20-encoding iRICE protocol. The upper panel shows iRICE maps fitted with WLLS and CWLLS,
followed by WLLS--CWLLS difference maps. The lower panel shows WLLS pre-fit constraint violations
and CWLLS post-fit violations. As in the healthy volunteer, CWLLS mitigates outlier values in
fluid-rich voxels, including the ventricles, cortical CSF interfaces, and the necrotic lesion core.
Lesion tissue and brain parenchyma appear less affected. However, difference maps show that CWLLS
removes implausible negative $\mathrm{MK}$, $C_{\mathrm{MD}}$ and $V_{\mathrm{MD}}$ values in normal-appearing
brain tissue in the right hemisphere superior to the lesion. WLLS constraint violations spatially
coincide with the outliers seen in the WLLS maps. A third of voxels entered the constrained
optimization. The absence of post-fit violations indicates successful enforcement of the
constraints. MD is reported in
$\mu\mathrm{m}^{2}/\mathrm{ms}$, and $V_{\mathrm{MD}}$ is reported in
$\mu\mathrm{m}^{4}/\mathrm{ms}^{2}$.
}
\label{fig:patient_maps}
\end{figure*}

Figure~\ref{fig:patient_maps} shows the estimator comparison and constraint diagnosis in the glioma patient, with WLLS and CWLLS applied to the 20-encoding iRICE protocol. WLLS and CWLLS show similar lesion and parenchymal contrast. However, CWLLS produced substantially fewer high-valued outliers in the ventricles and the necrotic lesion core, across all parameters. The spatial overlap between WLLS pre-fit violations, WLLS outliers, and difference-map patterns indicates that these outliers arise from unphysical fitting results. The $\mathrm{MK}$, $C_{\mathrm{MD}}$ and $V_{\mathrm{MD}}$ difference maps further show removal of a significant amount of erroneous negative values in normal-appearing brain tissue in the right hemisphere superior to the lesion. After CWLLS, post-fit violations were absent, indicating successful constraint enforcement.

The diagnostics showed that 32.7\% of voxels (13,247/40,474) lay outside at least one bound after the initial WLLS fit and were therefore passed to constrained optimization. For post hoc comparison, the QTI$\pm$ constant scalar cap was also evaluated and was violated in 1.7\% of voxels (687/40,474), whereas the new mean-dependent bound introduced here was violated in 15.4\% (6,249/40,474).

\begin{figure*}[!t]
\centering
\includegraphics[width=0.90\textwidth]{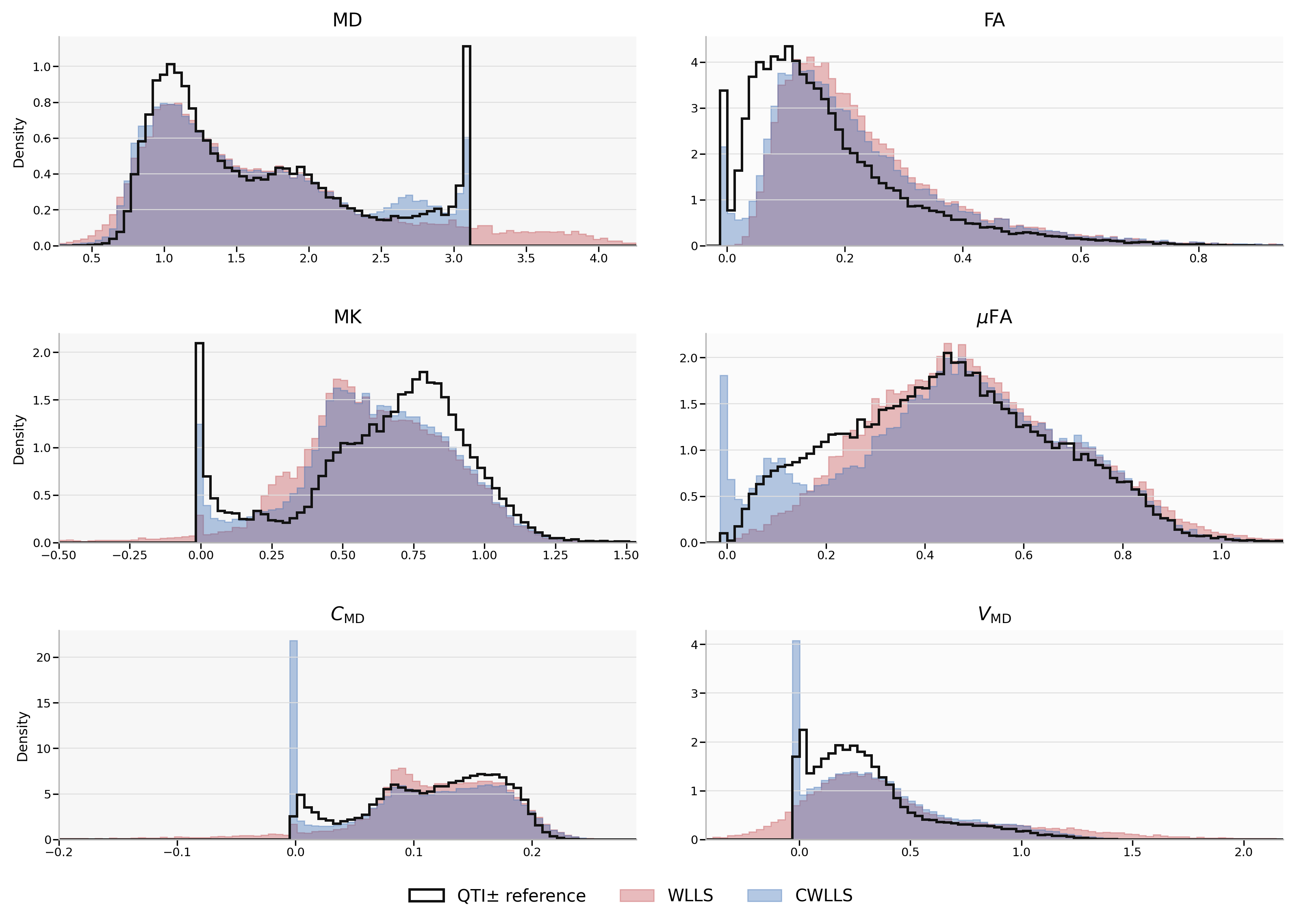}
\caption{
Whole-brain histogram comparison in the glioma patient acquired with the 1.5-min,
20-encoding iRICE protocol. Histograms show the distributions of invariant maps fitted with the 
54-encoding QTI
reference using QTI$\pm$ and with iRICE using WLLS and CWLLS. Purple regions indicate overlap
between the red WLLS and blue CWLLS histograms. Across metrics, WLLS and CWLLS showed overall
agreement with the reference distribution. Consistent with the spatial maps and
constraint-violation overlays in Figure~\ref{fig:patient_maps}, estimator differences were
concentrated in the distribution tails and near constraint boundaries. The constrained fit removed
MD values above the diffusivity limit, as expected, and mitigated implausible tails across the other
bounded metrics, including negative $\mathrm{MK}$, $C_{\mathrm{MD}}$ and $V_{\mathrm{MD}}$ values and $\mu$FA
values exceeding its normalized range of $[0,1]$. MD is reported in
$\mu\mathrm{m}^{2}/\mathrm{ms}$, and $V_{\mathrm{MD}}$ is reported in
$\mu\mathrm{m}^{4}/\mathrm{ms}^{2}$.
}
\label{fig:patient_wholebrain_histograms}
\end{figure*}

Figure~\ref{fig:patient_wholebrain_histograms} shows the corresponding whole-brain distributions. The dominant peaks of WLLS, CWLLS, and QTI$\pm$ were broadly aligned across the displayed invariants. Differences mainly appeared in the tails, where WLLS produced out-of-range values, including MD above the diffusivity limit, negative $C_{\mathrm{MD}}$ and $V_{\mathrm{MD}}$, and values above one for $\mu$FA. CWLLS truncated these tails and concentrated affected voxels near the corresponding bounds, consistent with the localized constraint effects seen in Figure~\ref{fig:patient_maps}.

\subsection{Resolution-dependent SNR behavior}

\begin{figure*}[!t]
\centering
\includegraphics[height=0.48\textheight,keepaspectratio]{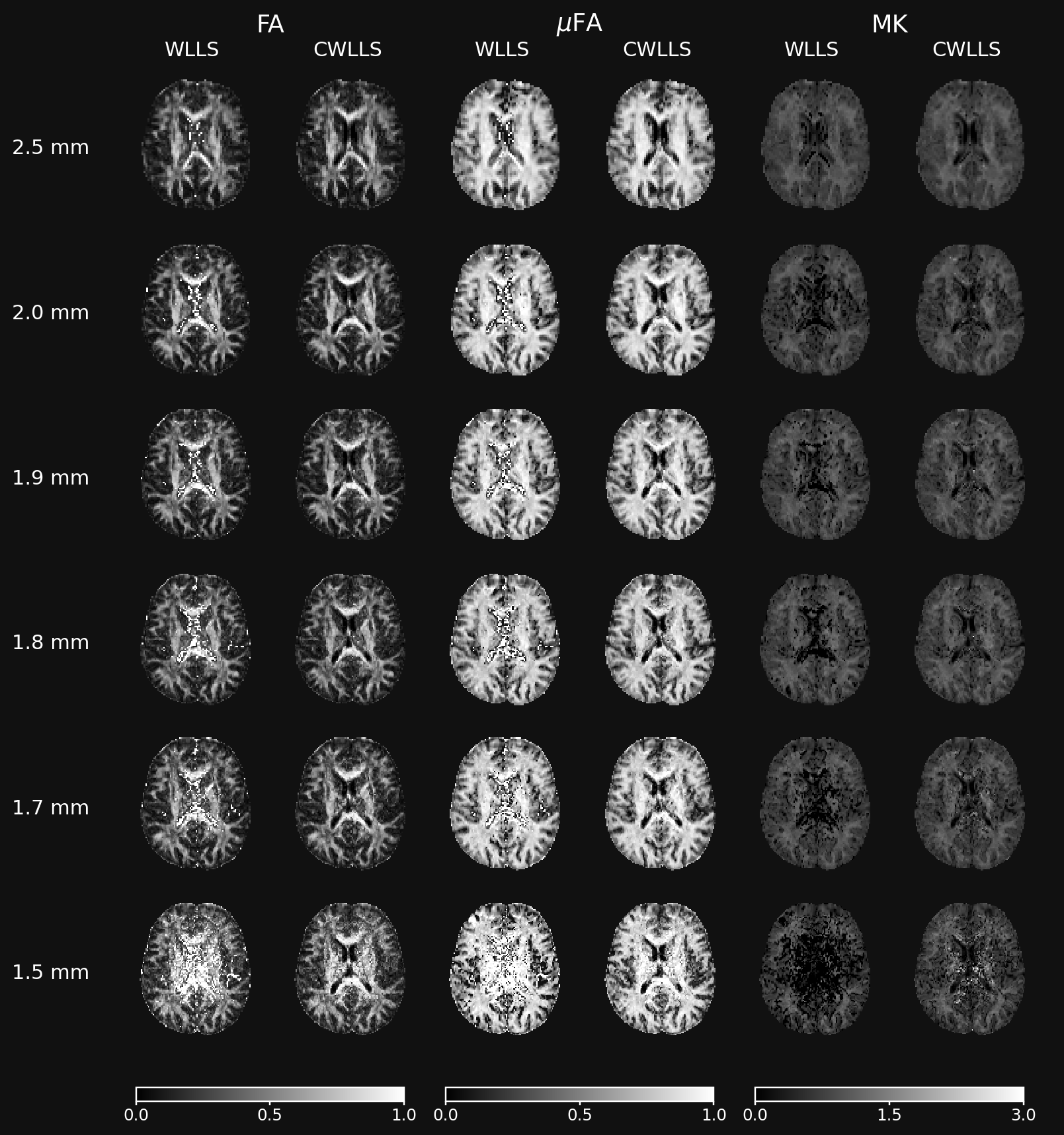}
\caption{
Resolution-dependent SNR experiment in a healthy volunteer acquired with the 1.5-min, 20-encoding
iRICE protocol. WLLS and CWLLS were compared across increasingly high isotropic spatial
resolutions, corresponding to reduced voxel volume and lower effective SNR. Maps are
shown for FA, $\mu$FA, and MK. From 2.5~mm to 1.5~mm isotropic resolution, voxel volume alone
reduces the expected SNR to approximately one-fifth, even before accounting for the TE increase from
84 to 100~ms. With increasing resolution, fitting artifacts first emerge in the image interior and
spread outward, consistent with higher receive sensitivity near the coil elements. These artifacts
often drive fitted values to the parameter bounds. CWLLS remains almost unaffected at 1.8~mm
isotropic resolution and retains more coherent anatomical structure beyond this point. At the
highest resolutions, both estimators deteriorate, but WLLS degrades more severely.
}
\label{fig:resolution_snr_robustness}
\end{figure*}

Figure~\ref{fig:resolution_snr_robustness} shows estimator behavior with increasing isotropic spatial resolution and decreasing effective SNR. Parameter maps retained recognizable anatomical structure and remained relatively smooth up to 1.8~mm isotropic voxel size. Across resolutions, CWLLS was visibly more stable than WLLS, with substantially fewer fitting outliers in the ventricles and central regions. At 1.5~mm, strong noise-related degradation became apparent in both estimators, although CWLLS retained greater map coherence.

\subsection{Computation time}

For the same healthy volunteer shown in Section~\ref{subsec:healthy_volunteer}, both constrained methods were benchmarked using the 54-encoding reference protocol. QTI$\pm$ required 72 minutes. In contrast, CWLLS required only 27 seconds, which amounts to a 160-fold acceleration.

\section{Discussion}

The one- to two-minute iRICE protocols introduced by Coelho et al. substantially reduce the acquisition-time barrier to clinical use of tensor-valued dMRI. Our main contribution to this line of work is the formulation of constraints directly in the reduced iRICE coefficient space, which stabilize fitting under the limited redundancy of these short protocols. The constraints acted mainly where WLLS became unreliable: CWLLS narrowed error distributions in the CSF-dominant simulation, reduced fit outliers in vivo, removed negative variance estimates in brain tissue, and cut off out-of-bounds histogram tails. The resolution experiment extended this pattern to lower-SNR conditions, where CWLLS retained more coherent maps than WLLS. At the highest resolution, both estimators degraded, indicating that the constraints improve robustness but cannot replace adequate SNR.

QTI+ and diffusivity-limited QTI$\pm$ established the value of physical constraints for full QTI \cite{herberthson2021qti,boito2023diffusivity}, but they operate on the complete mean, covariance, and moment tensors. iRICE instead retains only eight coefficients required for the targeted markers, compared with 27 parameters in full QTI, yielding a smaller constrained problem with fewer poorly conditioned terms to absorb noise. Combined with selective optimization of only those voxels that violated at least one bound after WLLS initialization, this yielded sub-30-second fits on standard PC hardware and a 160-fold runtime gain over QTI$\pm$, while preserving useful scalar-map quality from a minimal-time acquisition.

This computational gain matters clinically. A short acquisition loses much of its practical value if fitting still requires tens of minutes or remains confined to offline processing. By combining a 1.5-minute acquisition with near-instant constrained fitting, CWLLS and iRICE bring tensor-valued dMRI within the time demands of routine clinical workflows.

The constraint diagnostics identify the new mean-dependent bound as a key component of CWLLS. In the glioma patient dataset, this bound was violated in 15.4\% of voxels, compared with only 1.7\% for the QTI$\pm$ constant scalar cap \cite{boito2023diffusivity}. This indicates that the new bound is not merely a theoretical refinement. It changes the feasible parameter set in vivo and affects estimates that the looser QTI$\pm$ bound would not control.

Several limitations follow. First, physically plausible estimates do not imply accuracy. CWLLS can remove unphysical or extreme WLLS estimates, but estimates near active bounds may acquire bias. The controlled simulations showed this bias--variance trade-off most clearly. Second, the diffusivity limit was fixed. A bound set too low could over-constrain real high-diffusivity components, whereas a bound set too high would weaken the constraint. Additionally, our CWLLS constraints should not be interpreted as a change-of-basis translation of the full QTI$\pm$ constraint set. They enforce necessary conditions on the fitted iRICE coefficients, rather than the full-tensor constraints used in QTI$\pm$.

Future work could test sensitivity to the diffusivity limit and assess repeatability, reproducibility across scanners, and the diagnostic performance in larger patient cohorts. Finally, residual-based weighting may further improve robustness to motion and other artifacts. iRICE protocols, however, leave little redundancy, so aggressive outlier rejection could worsen conditioning. Still, carefully tuned weights remain worth testing, as suggested by recent work on constrained fitting in cardiac diffusion kurtosis imaging~\cite{coveney2026robust}.

\section{Conclusion}

In this work, we introduced CWLLS for iRICE and showed that this combination makes clinically practical tensor-valued dMRI much more realistic. It preserves the speed advantage of iRICE, constrains parameter combinations prone to unstable or physically invalid estimates, and cuts fitting time by orders of magnitude relative to full QTI$\pm$ in the present implementation. By enabling fast, constrained estimation of markers such as $\mu$FA, this approach may support broader clinical applications, where scan time, spatial resolution, estimation speed, and fit stability all matter.

\section*{Acknowledgements}
The work is part of the CZS Heidelberg Initiative for Model-Based AI (MBAI, P2021-02-001), which is funded by the Carl Zeiss Foundation. The authors gratefully acknowledge their support.

\section*{Data Availability Statement}

The imaging data are not publicly available because of data protection restrictions. The CWLLS fitting implementation is publicly available at \url{https://github.com/CollapsarEcho/iRICE-CWLLS}.

\section*{Supporting information}

The Supplementary Information is organized into three sections. Section 1 defines the RICE-derived scalar metrics used to compute the dMRI contrasts reported in the main text. Section 2 documents the CWLLS constraint set and derives the mean-dependent upper bounds and their scope. Section 3 presents healthy-volunteer maps obtained with the 20-encoding iRICE protocol, complementing the 54-encoding comparison in the main text.

\nocite{martin2020cnr,goedicke2024accelerated}
\bibliography{MRM-AMA}

\clearpage
\onecolumn
\setcounter{section}{0}
\setcounter{subsection}{0}
\setcounter{equation}{0}
\setcounter{figure}{0}
\setcounter{table}{0}
\renewcommand{\thesection}{\arabic{section}}
\renewcommand{\thesubsection}{\thesection.\arabic{subsection}}
\renewcommand{\theequation}{S\arabic{equation}}
\renewcommand{\thefigure}{S\arabic{figure}}
\renewcommand{\thetable}{S\arabic{table}}

\begin{center}
{\LARGE\bfseries Supplementary Information\\[0.4em]
\large Constrained estimation of rotational invariants of the 
cumulant expansion (RICE) for rapid tensor-valued diffusion MRI\par}
\vspace{1em}
{\normalsize Jinyang Yu$^{*}$, Oliver G\"odicke$^{*}$, Frederik B. Laun,
Obada T. Alhalabi, Iris A. Kohler, J\"urgen Hesser, Sandro M. Krieg, Bogdana Suchorska,
Heinz-Peter Schlemmer,
Mark E. Ladd, David Bonekamp, Johann M. E. Jende,
and Tristan A. Kuder\par}
\vspace{0.5em}
\small $^{*}$Jinyang Yu and Oliver G\"odicke contributed equally to this work.
\end{center}

\section*{Overview}

This Supplementary Information has three sections. Section 1 defines the
scalar invariants derived from the fitted iRICE coefficients. Section 2 documents
the CWLLS constraints on the mean diffusion tensor, $\Qzero$, and $\Tzero$,
and derives the mean-dependent upper bound and its scope. Section 3 presents two supplementary figures:
representative healthy-volunteer maps for the 20-encoding iRICE protocol and
pooled cohort-level histograms.

\section{Scalar metrics derived from fitted iRICE coefficients}

After fitting, the scalar maps used in this work were computed from the fitted
iRICE coefficients. The RICE expressions below follow Coelho
et al.~\cite{coelho2026geometry}, except for the normalized size-variance index
$C_{\mathrm{MD}}$, for which we use the definition of Westin
et al.~\cite{westin2016qspace}. The mean diffusivity was
\begin{equation}
\mathrm{MD}=\Dzero .
\label{supp:eq:md_rice}
\end{equation}
The fractional anisotropy of the mean diffusion tensor was
\begin{equation}
\mathrm{FA}
=
\left[
\frac{3\Dtwo^2}
{4\Dzero^2+2\Dtwo^2}
\right]^{1/2}.
\label{supp:eq:fa_rice}
\end{equation}
The isotropic diffusivity variance was
\begin{equation}
V_{\mathrm{MD}}=\Qzero ,
\label{supp:eq:vmd_rice}
\end{equation}
and the corresponding normalized size-variance index was
\begin{equation}
C_{\mathrm{MD}}
=
\frac{\Qzero}{\Qzero+\Dzero^2}.
\label{supp:eq:cmd_rice}
\end{equation}

Microscopic fractional anisotropy was then computed as
\begin{equation}
\mu\mathrm{FA}
=
\left[
\frac{15\Tzero+3\Dtwo^2}
{10\Tzero+2\Dtwo^2+4\Dzero^2}
\right]^{1/2}.
\label{supp:eq:ufa_rice}
\end{equation}
Finally, the trace-based mean-kurtosis contrast, denoted by
$\mathrm{MK}$ here~\cite{coelho2026geometry}, was computed from the scalar covariance components as
\begin{equation}
\mathrm{MK}
=
\frac{3(\Qzero+\Tzero)}{\Dzero^2}.
\label{supp:eq:mk_rice}
\end{equation}

\section{CWLLS constraint set}

Throughout this section, $\Dmicro$ denotes a microscopic diffusion tensor
drawn from the DTD within one voxel, and $\E_{\mathrm{DTD}}$ denotes
expectation over that distribution. Roman symbols denote the corresponding
voxel-wise quantities; in particular,
$\Dmean\equiv\E_{\mathrm{DTD}}[\Dmicro]$. Thus, this expectation averages
microscopic tensors within a voxel.

\subsection{Mean diffusion tensor constraints}

The mean-tensor constraints are imposed after reconstructing the Cartesian
mean diffusion tensor from the fitted iRICE coefficients. In the coefficient
convention used by the implementation, the mean-diffusion coefficients are
stored in the order
\begin{equation}
\boldsymbol{d}_{\mathrm{STF}}
=
\left(
\Dzero,\,
\mathrm{D}^{22},\,
\mathrm{D}^{21},\,
\mathrm{D}^{20},\,
\mathrm{D}^{2,-1},\,
\mathrm{D}^{2,-2}
\right)^{\mathrm T}.
\label{supp:eq:stf_mean_vector}
\end{equation}
The corresponding Cartesian mean diffusion tensor is
\begin{equation}
\Dmean(\boldsymbol{d}_{\mathrm{STF}})
=
\begin{pmatrix}
\Dzero-\frac{1}{2}\mathrm{D}^{20}
+\frac{\sqrt{3}}{2}\mathrm{D}^{2,-2}
&
\frac{\sqrt{3}}{2}\mathrm{D}^{22}
&
\frac{\sqrt{3}}{2}\mathrm{D}^{2,-1}
\\
\frac{\sqrt{3}}{2}\mathrm{D}^{22}
&
\Dzero-\frac{1}{2}\mathrm{D}^{20}
-\frac{\sqrt{3}}{2}\mathrm{D}^{2,-2}
&
\frac{\sqrt{3}}{2}\mathrm{D}^{21}
\\
\frac{\sqrt{3}}{2}\mathrm{D}^{2,-1}
&
\frac{\sqrt{3}}{2}\mathrm{D}^{21}
&
\Dzero+\mathrm{D}^{20}
\end{pmatrix}.
\label{supp:eq:rice_to_cartesian_mean_tensor}
\end{equation}
For the optimizer, the same tensor is represented in Kelvin form:
\begin{equation}
\operatorname{vec}_{\mathrm K}(\Dmean)
=
\left(
\mathrm{D}_{xx},\,
\mathrm{D}_{yy},\,
\mathrm{D}_{zz},\,
\sqrt{2}\mathrm{D}_{xy},\,
\sqrt{2}\mathrm{D}_{xz},\,
\sqrt{2}\mathrm{D}_{yz}
\right)^{\mathrm T}.
\label{supp:eq:kelvin_mean_vector}
\end{equation}
The transformation is the linear map
\begin{equation}
\operatorname{vec}_{\mathrm K}(\Dmean)
=
\underbrace{
\begin{pmatrix}
1&0&0&-\frac{1}{2}&0& \frac{\sqrt{3}}{2}\\
1&0&0&-\frac{1}{2}&0&-\frac{\sqrt{3}}{2}\\
1&0&0&1&0&0\\
0&\sqrt{\frac{3}{2}}&0&0&0&0\\
0&0&0&0&\sqrt{\frac{3}{2}}&0\\
0&0&\sqrt{\frac{3}{2}}&0&0&0
\end{pmatrix}
}_{\boldsymbol{K}_{\mathrm{STF}\to\mathrm K}}
\boldsymbol{d}_{\mathrm{STF}}.
\label{supp:eq:stf_to_kelvin_mean_tensor}
\end{equation}

The semidefinite constraints apply to the reconstructed $3\times3$ tensor,
not to individual entries of its Kelvin vector
~\cite{herberthson2021qti,boito2023diffusivity}:
\begin{equation}
\Dmean\succeq 0,
\qquad
\Dmax\Ibold-\Dmean\succeq 0.
\label{supp:eq:mean_tensor_constraints_complete}
\end{equation}
These two matrix inequalities require all eigenvalues of the voxel-wise mean
diffusion tensor to lie in $[0,\Dmax]$, with
$\Dmax=3.075~\mu\mathrm{m}^2/\mathrm{ms}$.

\subsection{Non-negativity of \texorpdfstring{$\Qzero$ and $\Tzero$}{Q0 and T0}}

The scalar RICE covariance invariants $\Qzero$ and $\Tzero$ represent the
isotropic diffusivity variance and the rotationally averaged covariance of
the microscopic anisotropic tensor components, respectively. Both are required to be:
\begin{equation}
\Qzero\geq0,
\qquad
\Tzero\geq0.
\label{supp:eq:scalar_nonnegativity}
\end{equation}
These are linear constraints in the fitted iRICE parameterization.

\subsection{Mean-dependent upper-bound constraint}

CWLLS uses the following mean-dependent bound on the total directional
diffusivity variance:
\begin{equation}
\Qzero+\Tzero+\frac{1}{5}\Dtwo^2
\leq \Dzero(\Dmax-\Dzero).
\label{supp:eq:total_bound_overview}
\end{equation}
The following subsections derive this constraint.
\subsubsection{Bounded directional diffusivity}

Let $\Dmicro$ be drawn from the voxel DTD and let $\ubar$ be an independent,
uniformly distributed direction on $\sphere$. For every microscopic tensor,
assume the diffusivity limit of Boito et al.~\cite{boito2023diffusivity}:
\begin{equation}
0 \preceq \Dmicro \preceq \Dmax\Ibold .
\label{supp:eq:microscopic_bound}
\end{equation}
For a unit direction $\ubar$, define
\begin{equation}
X=\ubar^{\mathrm T}\Dmicro\ubar .
\label{supp:eq:x_definition}
\end{equation}
Equation~\eqref{supp:eq:microscopic_bound} implies
\begin{equation}
0\leq X\leq\Dmax .
\label{supp:eq:x_bounded}
\end{equation}

The diffusivity-limited QTI$\pm$ appendix uses this interval to derive the
fixed Popoviciu variance bound~\cite{boito2023diffusivity}
\begin{equation}
\Var_{\mathrm{DTD},\ubar}(X)\leq\frac{\Dmax^2}{4}.
\label{supp:eq:popoviciu_bound}
\end{equation}
Because the mean is also known, the Bhatia--Davis inequality provides the
tighter, mean-dependent bound~\cite{bhatia2000better}. For
$\mu=\E_{\mathrm{DTD},\ubar}[X]$,
\begin{equation}
\Var_{\mathrm{DTD},\ubar}(X)\leq \mu(\Dmax-\mu).
\label{supp:eq:bhatia_davis}
\end{equation}
The proof is immediate~\cite{bhatia2000better}:
\begin{equation}
0\leq X(\Dmax-X)
\quad\Rightarrow\quad
\E[X^2]\leq \Dmax\mu,
\quad\Rightarrow\quad
\Var(X)=\E[X^2]-\mu^2\leq \mu(\Dmax-\mu).
\label{supp:eq:bhatia_davis_proof}
\end{equation}

It remains to evaluate $\mu$ in RICE notation. Following the
Racah-normalized spherical-tensor decomposition~\cite{coelho2026geometry}, with
$Y^{2m}(\ubar)=\mathcal{Y}^{2m}_{ij}u_i u_j$, the directional diffusivity of
one microscopic tensor is
\begin{equation}
X(\ubar)
=
\Dmicrozero
+
\sum_{m=-2}^{2}\Dmicrotwom Y^{2m}(\ubar),
\qquad
\Dmicrozero=\frac{1}{3}\operatorname{tr}\Dmicro.
\label{supp:eq:x_stf}
\end{equation}
Here, $\Dmicrozero$ and $\Dmicrotwom$ are the expansion coefficients of that
single microscopic tensor. Their DTD averages are the voxel-wise RICE
coefficients $\Dzero$ and $\Dtwom$, respectively. Because the $\ell=2$ terms
have zero directional mean,
\begin{equation}
\mu
=\E_{\mathrm{DTD},\ubar}[X]
=\E_{\mathrm{DTD}}[\Dmicrozero]
=\Dzero .
\label{supp:eq:x_mean}
\end{equation}
Substitution into Eq.~\eqref{supp:eq:bhatia_davis} gives
\begin{equation}
\Var_{\mathrm{DTD},\ubar}(X)
\leq \Dzero(\Dmax-\Dzero).
\label{supp:eq:right_hand_side}
\end{equation}
This bound is never larger than the fixed bound in
Eq.~\eqref{supp:eq:popoviciu_bound}, because
\begin{equation}
\Dzero(\Dmax-\Dzero)
= \frac{\Dmax^2}{4}
-\left(\Dzero-\frac{\Dmax}{2}\right)^2
\leq \frac{\Dmax^2}{4}.
\label{supp:eq:constant_bound_relation}
\end{equation}
The mean-dependent and constant upper bounds coincide only at $\Dzero=\Dmax/2$.

\subsubsection{Directional variance in RICE invariants}

For a function of direction $f$, define the normalized spherical average as
\begin{equation}
\E_{\ubar}[f(\ubar)]
=
\frac{1}{4\pi}\int_{\sphere}f(\ubar)\,d\Omega .
\label{supp:eq:directional_average}
\end{equation}

Under the Racah normalization used by Coelho et al.~\cite{coelho2026geometry},
\begin{equation}
\E_{\ubar}\!\left[Y^{2m}(\ubar)\right]=0,
\qquad
\E_{\ubar}\!\left[
Y^{2m}(\ubar)Y^{2n*}(\ubar)
\right]
=
\frac{1}{5}\delta_{mn}.
\label{supp:eq:spherical_orthogonality}
\end{equation}

The second identity, and hence the factor $1/5$, is the $\ell=2$ spherical
average corresponding to Supplementary Eq.~S28 ~\cite{coelho2026geometry} under the same Racah normalization. Expanding
Eq.~\eqref{supp:eq:x_stf} and using Eq.~\eqref{supp:eq:spherical_orthogonality}, the
cross terms vanish after spherical averaging, giving
\begin{equation}
\E_{\mathrm{DTD},\ubar}[X^2]
=
\E_{\mathrm{DTD}}[\Dmicrozero^2]
+\frac{1}{5}\sum_{m=-2}^{2}
\E_{\mathrm{DTD}}\left[\Dmicrotwom\Dmicrotwomstar\right].
\label{supp:eq:x_second_moment}
\end{equation}
Since $\E_{\mathrm{DTD},\ubar}[X]=\Dzero$, the corresponding variance is
\begin{align}
\Var_{\mathrm{DTD},\ubar}(X)
&=\E_{\mathrm{DTD},\ubar}[X^2]-\Dzero^2 \notag\\
&=\left(\E_{\mathrm{DTD}}[\Dmicrozero^2]-\Dzero^2\right)
+\frac{1}{5}\sum_{m=-2}^{2}
\E_{\mathrm{DTD}}\left[\Dmicrotwom\Dmicrotwomstar\right].
\label{supp:eq:x_variance_raw}
\end{align}
The first term is the RICE isotropic variance defined in Eq.~8a of Coelho
et al.~\cite{coelho2026geometry}:
\begin{equation}
\Qzero
= \Var_{\mathrm{DTD}}(\Dmicrozero)
=\E_{\mathrm{DTD}}[\Dmicrozero^2]-\Dzero^2.
\label{supp:eq:q0_definition}
\end{equation}

The last term in Eq.~\eqref{supp:eq:x_variance_raw} is the DTD average of the
squared microscopic anisotropy norm,
\begin{equation}
\E_{\mathrm{DTD}}[\Dmicrotwo^2]
\equiv
\sum_{m=-2}^{2}
\E_{\mathrm{DTD}}\left[\Dmicrotwom\Dmicrotwomstar\right].
\label{supp:eq:microscopic_d2_moment}
\end{equation}
Equation~29 of Coelho et al.~\cite{coelho2026geometry} gives directly
\begin{equation}
\E_{\mathrm{DTD}}[\Dmicrotwo^2]
=5\Tzero+\Dtwo^2.
\label{supp:eq:coelho_d2_moment}
\end{equation}
Therefore, Eq.~\eqref{supp:eq:x_variance_raw} becomes
\begin{equation}
\Var_{\mathrm{DTD},\ubar}(X)
= \Qzero + \Tzero + \frac{1}{5}\Dtwo^2 .
\label{supp:eq:x_variance_rice}
\end{equation}

In detail, with
$\Dtwom=\E_{\mathrm{DTD}}[\Dmicrotwom]$, the last term in
Eq.~\eqref{supp:eq:x_variance_raw} separates into its second cumulant and squared
mean:
\begin{align}
\E_{\mathrm{DTD}}[\Dmicrotwo^2]
&=
\sum_{m=-2}^{2}
\Cov_{\mathrm{DTD}}\left(\Dmicrotwom,\Dmicrotwomstar\right)
+\sum_{m=-2}^{2}\Dtwom\Dtwomstar.
\label{supp:eq:raw_second_split}
\end{align}
The covariance is the second cumulant denoted by double brackets in Coelho et
al. Using their Eqs.~60 and 25, respectively,
\begin{align}
\Tzero
&=
\frac{1}{5}\sum_{m=-2}^{2}
\left\langle\!\left\langle
\Dmicrotwom\Dmicrotwomstar
\right\rangle\!\right\rangle_{\mathrm{DTD}}
\notag\\
&=
\frac{1}{5}\sum_{m=-2}^{2}
\Cov_{\mathrm{DTD}}(\Dmicrotwom,\Dmicrotwomstar),
\label{supp:eq:t0_definition}\\
\Dtwo^2
&=\sum_{m=-2}^{2}\Dtwom\Dtwomstar.
\label{supp:eq:d2_norm}
\end{align}

This is the same diffusional-variance decomposition used in the DIVIDE/QTI
literature, where the total diffusional variance is written as an isotropic
part plus an anisotropic part, $V_{\mathrm{I}}+V_{\mathrm{A}}$
\cite{szczepankiewicz2016divide,westin2016qspace}. In the present RICE notation~\cite{coelho2026geometry},
\begin{equation}
V_{\mathrm{I}}=\Qzero,
\qquad
V_{\mathrm{A}}=\Tzero+\frac{1}{5}\Dtwo^2 .
\label{supp:eq:divide_variance_identification}
\end{equation}

\subsubsection{Final constraint and scope}

Combining Eq.~\eqref{supp:eq:right_hand_side} with Eq.~\eqref{supp:eq:x_variance_rice} gives
the bound in Eq.~\eqref{supp:eq:total_bound_overview}. In the form supplied to the
convex optimizer, it is
\begin{equation}
\Qzero+\Tzero+\frac{1}{5}\Dtwo^2
+\left(\Dzero-\frac{\Dmax}{2}\right)^2
\leq \frac{\Dmax^2}{4}.
\label{supp:eq:total_variance_convex_form}
\end{equation}

Because $\Qzero$ and $\Tzero$ enter affinely in the fitted iRICE
coefficients, whereas
$\Dtwo^2$ and
$\left(\Dzero-\Dmax/2\right)^2$ are convex quadratic functions, the
left-hand side of Eq.~\eqref{supp:eq:total_variance_convex_form} is convex.
The corresponding sublevel-set constraint is therefore convex.
Note that Eq.~\eqref{supp:eq:total_bound_overview} is a necessary scalar consequence of
Eq.~\eqref{supp:eq:microscopic_bound} after averaging over directions. It does not
replace the full QTI+ and QTI$\pm$ constraint sets. In addition to constraints
on the complete mean tensor, full QTI+ imposes positivity conditions on the
$6\times6$ representation of the covariance tensor and on the fourth-order
second-moment tensor. QTI$\pm$ further imposes upper-diffusivity conditions on
these full tensor objects~\cite{herberthson2021qti,boito2023diffusivity}.

\clearpage
\section{Healthy-volunteer comparisons using the 20-encoding iRICE protocol}

\begin{figure}[htbp]
\centering
\includegraphics[width=\textwidth]{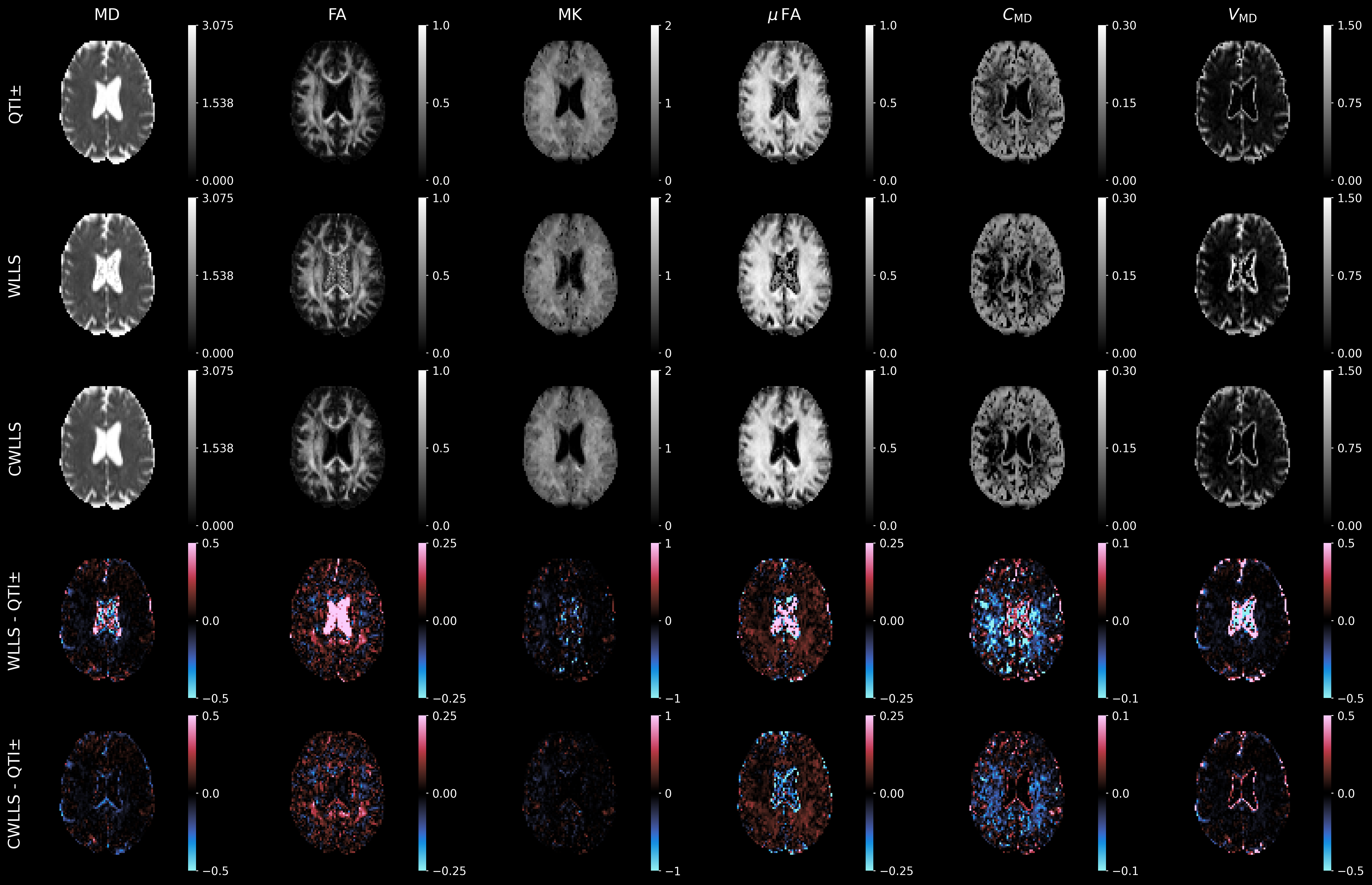}
\caption{Healthy-volunteer maps for the 1.5-minute, 20-encoding iRICE
protocol. Columns show MD, FA, MK, $\mu\mathrm{FA}$, $C_{\mathrm{MD}}$, and
$V_{\mathrm{MD}}$. Rows show the 54-encoding QTI$\pm$ reference, WLLS and
CWLLS fitted to the same 20-encoding data, and their signed differences from
QTI$\pm$. CWLLS generally showed closer agreement with the QTI$\pm$ reference
than WLLS. Because the reference and iRICE estimates were derived from
different acquisition protocols, these differences reflect both estimator
effects and differences in acquired information content. MD is reported in $\mu\mathrm{m}^{2}/\mathrm{ms}$,
and $V_{\mathrm{MD}}$ is reported in
$(\mu\mathrm{m}^{2}/\mathrm{ms})^{2}$.}
\label{supp:fig:healthy_irice20_maps}
\end{figure}

Figure~\ref{supp:fig:healthy_irice20_maps} complements the controlled 54-encoding
comparison in the main text by showing WLLS and CWLLS fitted with the 
20-encoding iRICE protocol.

\clearpage

\begin{figure}[htbp]
\centering
\includegraphics[width=\textwidth]{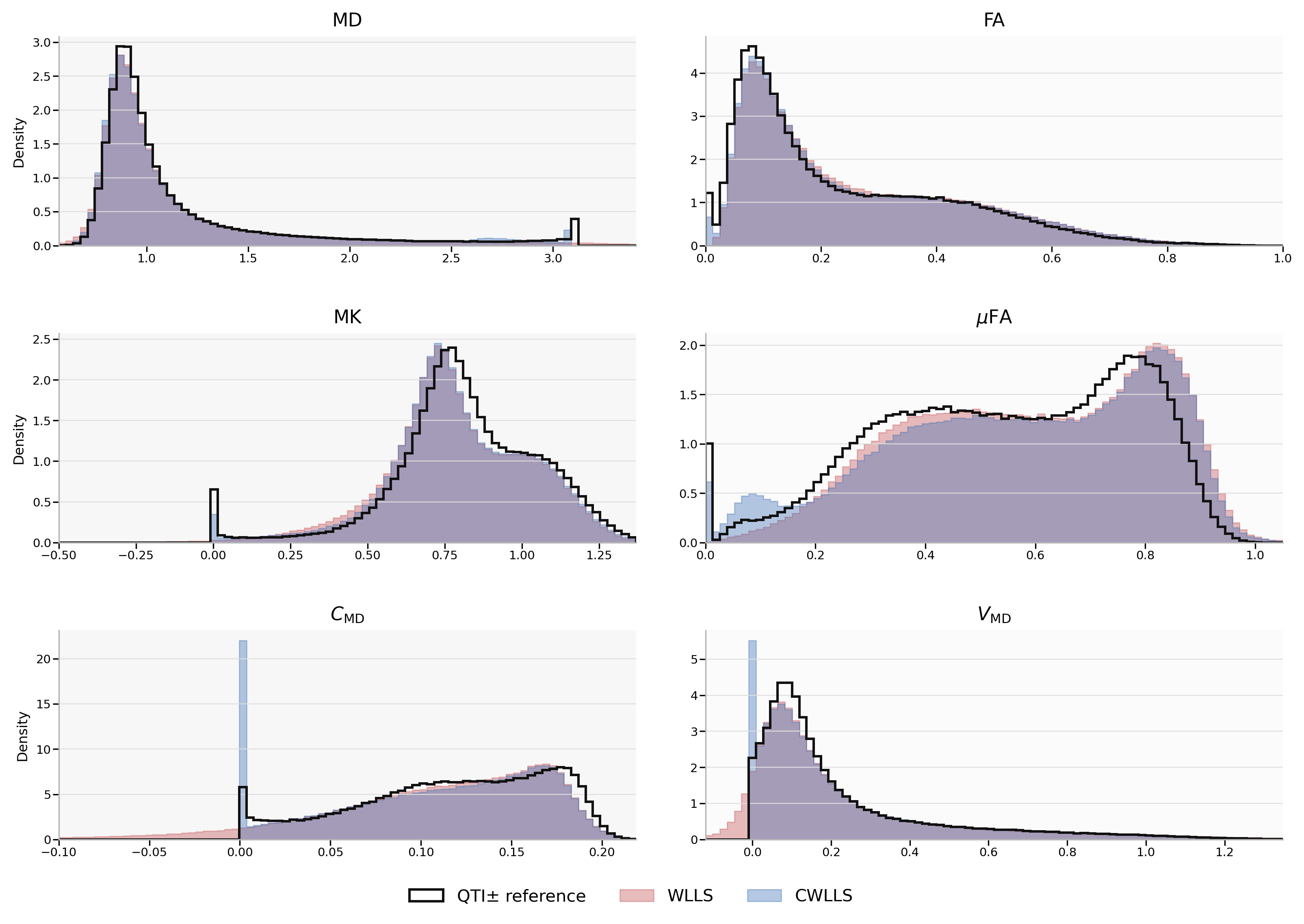}
\caption{Pooled healthy-volunteer histogram comparison. 
Density histograms show MD, FA, MK, $\mu\mathrm{FA}$,
$C_{\mathrm{MD}}$, and $V_{\mathrm{MD}}$ for the central 12 slices in
14 volunteers. The QTI$\pm$ reference used 54 encodings. WLLS and CWLLS used
the 20-encoding iRICE protocol. Values were pooled voxel-wise across
the cohort. Across metrics, both iRICE estimators showed overall agreement
with the reference distributions. Differences were concentrated in the
distribution tails, where CWLLS suppressed the unphysical values observed
with WLLS, including out-of-range MD and $\mu\mathrm{FA}$ estimates and
negative $C_{\mathrm{MD}}$ and $V_{\mathrm{MD}}$ values. Display windows limit
the shown tails but do not alter the source values. MD is reported in
$\mu\mathrm{m}^{2}/\mathrm{ms}$, and $V_{\mathrm{MD}}$ is reported in
$(\mu\mathrm{m}^{2}/\mathrm{ms})^{2}$.}
\label{supp:fig:healthy_pooled_histograms}
\end{figure}

Analogous to the single-patient whole-brain histogram comparison in Figure~4
of the main text, Figure~\ref{supp:fig:healthy_pooled_histograms} summarizes
estimator distributions across the 14 healthy-volunteer cohort.

\clearpage

\end{document}